\documentclass[12pt]{article}
\usepackage[latin9]{inputenc}
\usepackage{geometry}
\usepackage{float}
\usepackage{mathtools}
\usepackage{amsmath}
\usepackage{amsthm}
\usepackage{amssymb}
\usepackage{graphicx}
\usepackage{setspace}
\usepackage[authoryear]{natbib}
\usepackage{latexsym,amsfonts,bm}
\usepackage{rotating,threeparttable}
\usepackage{subfigure,algorithmicx}
\usepackage{algpseudocode}
\usepackage{lscape}
\usepackage[pagewise,mathlines]{lineno}
\usepackage{appendix}

\providecommand{\tabularnewline}{\\}

\theoremstyle{plain}

  \theoremstyle{plain}

\newcommand*{\patchAmsMathEnvironmentForLineno}[1]{%
      \expandafter\let\csname old#1\expandafter\endcsname\csname #1\endcsname
      \expandafter\let\csname oldend#1\expandafter\endcsname\csname end#1\endcsname
      \renewenvironment{#1}%
         {\linenomath\csname old#1\endcsname}%
         {\csname oldend#1\endcsname\endlinenomath}}%
    \newcommand*{\patchBothAmsMathEnvironmentsForLineno}[1]{%
      \patchAmsMathEnvironmentForLineno{#1}%
      \patchAmsMathEnvironmentForLineno{#1*}}%
    \AtBeginDocument{%
    \patchBothAmsMathEnvironmentsForLineno{equation}%
    \patchBothAmsMathEnvironmentsForLineno{align}%
    \patchBothAmsMathEnvironmentsForLineno{flalign}%
    \patchBothAmsMathEnvironmentsForLineno{alignat}%
    \patchBothAmsMathEnvironmentsForLineno{gather}%
    \patchBothAmsMathEnvironmentsForLineno{multline}%
    }
\include{def}
\def\dispmuskip{\thinmuskip= 3mu plus 0mu minus 2mu \medmuskip=  4mu plus 2mu minus 2mu \thickmuskip=5mu plus 5mu minus 2mu}
\def\textmuskip{\thinmuskip= 0mu                    \medmuskip=  1mu plus 1mu minus 1mu \thickmuskip=2mu plus 3mu minus 1mu}
\def\beq{\dispmuskip\begin{equation}}    \def\eeq{\end{equation}\textmuskip}
\def\beqn{\dispmuskip\begin{displaymath}}\def\eeqn{\end{displaymath}\textmuskip}
\def\bea{\dispmuskip\begin{eqnarray}}    \def\eea{\end{eqnarray}\textmuskip}
\def\bean{\dispmuskip\begin{eqnarray*}}  \def\eean{\end{eqnarray*}\textmuskip}

\def\paradot#1{\vspace{1.3ex plus 0.7ex minus 0.5ex}\noindent{\bf\boldmath{#1.}}}

\newtheorem{algorithm}{Algorithm}

\newcommand{\KL}{Kullback-Leibler}

\newcommand{\diag}{\text{diag}}

\newcommand{\eps}{\epsilon}
\newcommand{\wh}{\widehat}

\def\E{{\mathbb E}}                         
\def\V{{\mathbb V}}
\def\P{{\rm P}}                         

\def\a{\alpha}

\def\s{\sigma}

\def\Sig{\Sigma}
\def\t{\theta}

\def\l{\lambda}

\def\N{{\cal N}}

\def\tr{\text{\rm tr}}

\def\cov{\text{\rm cov}}

\def\LB{\text{\rm LB}}

\def\KL{\text{\rm KL}}
\def\diag{\text{\rm diag}}

\makeatother
\providecommand{\propositionname}{Proposition}
\providecommand{\theoremname}{Theorem}

\begin{document}

\title{Fast Inference for Intractable Likelihood Problems using Variational Bayes}
\author{David Gunawan, Minh-Ngoc Tran and Robert Kohn\thanks{Gunawan and Kohn are with 
School of Economics, University of New South Wales Business School.
Tran is with Discipline of Business Analytics, University of Sydney Business School. Email: minh-ngoc.tran@sydney.edu.au}}
\maketitle

\begin{abstract}
Variational Bayes (VB) is a popular estimation method for Bayesian
inference. However, most existing VB algorithms are restricted to cases where
the likelihood is tractable, which precludes their use in many
important situations. \citet{Tran:2015} extend the scope of application
of VB to cases where the \textit{likelihood} is intractable but can
be estimated unbiasedly, and name the method Variational Bayes
with Intractable Likelihood (VBIL). This paper presents a version
of VBIL, named Variational Bayes with Intractable Log-Likelihood (VBILL),
that is useful for cases as Big Data and Big Panel Data models,
where unbiased estimators of the \textit{gradient} of the log-likelihood are
available. We demonstrate that such estimators can be easily obtained in many Big Data applications.
The proposed method is exact in the sense that,
apart from an extra Monte Carlo error which can be controlled, 
it is able to produce estimators as if the true likelihood, or full-data likelihood, is used. 
In particular, we develop a computationally efficient approach, based
on data subsampling and the MapReduce programming technique, for analyzing
massive datasets which cannot fit into the memory of a single desktop PC.
We illustrate the method using several simulated datasets and a big real dataset based on the arrival time status of U. S. airlines. 

\paradot{Keywords} Pseudo Marginal Metropolis-Hastings,
Big Data, Panel Data, Difference Estimator.
\end{abstract}

\section{Introduction}

Given an observed dataset $y$ and a statistical model with a vector
of unknown parameters $\theta$, a major aim of statistics is to carry
out inference about $\theta$, i.e., estimate the underlying $\theta$
that generated $y$ and assess the associated uncertainty. The likelihood
function $p(y|\theta)$, which is the density of the data $y$ conditional
on the postulated model and the parameter vector $\theta$, is the
basis of Bayesian methods, such as Markov chain Monte Carlo
(MCMC) and Variational Bayes (VB). These methods require exact evaluation of the likelihood $p(y|\theta)$
at each value of $\theta$. In many modern statistical applications,
however, the likelihood function, and thus the log-likelihood function, is either analytically intractable
or computationally intractable, making it difficult to use likelihood-based methods.

An important situation in which the log-likelihood is computationally
intractable is Big Data \citep{Bardenet2015,Quiroz:2015a}, where the log-likelihood function, under
the independence assumption and without random effects, is a sum of a very large number of terms
and thus too expensive to compute. Large panel data models \citep{Fitzmaurice:2011}
are another example where the log-likelihood is both analytically
and computationally intractable as it is a sum of many terms, each being the log of an integral over the random effects and cannot
be computed analytically.

There are several methods in the literature that work with an intractable
likelihood. A remarkable approach is the pseudo-marginal Metropolis-Hastings
(PMMH) algorithm \citep{Andrieu:2009}, which replaces the intractable likelihood
in the Metropolis-Hastings ratio by its non-negative unbiased estimator.
An attractive property of the PMMH approach is that it is exact
in the sense that it is still able to generate samples from the posterior, as if the true likelihood was used.
Similarly to standard Metropolis-Hastings algorithms, PMMH is extremely flexible. However,
this method is highly sensitive to the variance of the likelihood
estimator. The chain might get stuck and mix poorly if the likelihood
estimates are highly variable \citep{Flury:2011}. This is because
the asymptotic variance of PMMH estimators increases exponentially
with the variance of the log of the estimator of the likelihood \citep{Pitt:2012}, which in turn increases linearly with the sample size. 
Therefore the PMMH method can be computationally expensive making it unsuitable
for Big Data applications.

VB is a computationally efficient alternative to MCMC \citep{Attias:1999,Bishop:2006}.
However, most existing VB algorithms are restricted to cases where
the likelihood is tractable, which precludes the use of VB in many
interesting models. \citet{Tran:2015} extend the scope of application
of VB to cases where the likelihood is intractable but can be estimated
unbiasedly, and name the method Variational Bayes with Intractable
Likelihood (VBIL). Their method works with non-negative unbiased
estimators of the {\em likelihood}, and is useful when it is convenient 
to obtain unbiased estimates of the likelihood such as in state space models. This paper presents a version of VBIL, called the
Variational Bayes with Intractable Log-Likelihood (VBILL),
that requires unbiased estimates of the {\em gradient} of the log-likelihood.
It turns out that, in cases of Big Data both with and without random effects,
it is easy and computationally efficient to obtain unbiased estimates of the gradient of the log-likelihood using data subsampling. 
Consider the case of Big Data without random effects,
where the log-likelihood function is a sum of many terms, with each computed analytically.
Then the gradient of the log-likelihood is a sum of tractable terms, and this sum can be estimated unbiasedly using data subsampling.
In the case of Big Panel Data, each log-likelihood contribution is the log of an intractable integral and thus can no longer be computed analytically.
However, we are still able to obtain an unbiased estimate of the gradient of the log-likelihood 
using subsampling and Fisher's identity (see Section \ref{sec:VBILL-Big Panel}).
Although data subsampling has now been used in Big Data situations with a very large number of independent observations,
we are not aware of any efficient estimation methods for Big Panel Data models.
One of the main goals of our paper is to fill this gap.

This paper makes two important improvements to
VBIL that greatly enhance its performance. 
The first is that we take into account the information of the gradient of the log-likelihood,
which helps the stochastic optimization procedure more stable and converge faster.
The second is that we now aim to minimize the same Kullback-Leibler divergence as that targeted when the likelihood is tractable. 
That is, VBILL is exact in the sense that, apart from an extra Monte Carlo error which can be controlled, 
it is able to produce estimators as if the true likelihood or full-data likelihood were used. 
The VBIL approach of \cite{Tran:2015} is exact in this sense only under the condition that the variance of the likelihood estimator is constant.

Our paper also uses the MapReduce programming technique
and develops a computationally efficient approach for analyzing massive datasets which
do not fit into the memory of a single desktop PC. The implementation of MapReduce
uses the divide and combine idea where the data is divided into small
chunks, each chunk is processed separately and the chunk-based results
are then combined to construct the final estimates. Under some regularity
conditions, \citet{Battey:2015} show that the information loss due
to the divide and combine procedure is asymptotically negligible when
the full sample size grows, as long as the number of chunks is not
too large. In finite-sample settings, however, the resulting estimators
are sensitive to how the data are divided. It is important to note
that our final estimator is mathematically justified and independent
of the data chunking, as we use the divide and combine procedure mainly
to obtain an unbiased estimator of the gradient of log-likelihood
for the VBILL algorithm.

The paper is organized as follows.
Section \ref{sec:VBILL} describes the VBILL approach
and its applications to Big Data problems using data subsampling.
Section \ref{sec:Experimental studies} presents empirical studies
and Section \ref{sec:Conclusions} concludes.
The appendix presents technical details.

\section{Variational Bayes with Intractable Log-Likelihood}\label{sec:VBILL}

Let $p(\theta)$ be the prior, $L(\t):=p(y|\theta)$ the likelihood and $\pi(\theta)\propto p(\theta)L(\t)$ 
the posterior distribution of $\theta$. Let $\ell(\t):=\log L(\t)$.
Our paper, with a small abuse of notation, uses the same notation for the probability distribution and its density function.		
VB approximates the posterior distribution of $\theta$ by a probability distribution $q_{\lambda}(\theta)$ within some parametric class of distributions such as
an exponential family, with parameter $\lambda$ chosen to minimize
the Kullback-Leibler divergence between $q_{\lambda}(\theta)$
and $\pi(\theta)$, 
\[
\KL\left(\lambda\right)=\KL(q_{\lambda}\|\pi):=\int q_{\lambda}(\theta)\log\frac{q_{\lambda}(\theta)}{\pi(\theta)}d\theta.
\]
Minimizing this divergence is equivalent to maximizing the lower bound
\[\LB(\l)=\int q_\l(\t)\log\frac{p(\t)L(\t)}{q_\l(\t)}d\t=A(\l)+\int q_\l(\t)\ell(\t)d\t,\]
with $A(\l)=\int q_\l(\t)\log\frac{p(\t)}{q_\l(\t)}d\t$.
Often $A(\l)$ can be computed analytically.
Suppose that $\theta\sim q_\l(\t)$ can be represented as a deterministic function of a random vector $\eps$ whose distribution is independent of $\lambda$. 
More precisely, let $g(\cdot,\cdot)$ be a function such that $\t=g(\l,\eps)\sim q_\l(\t)$, where $\eps\sim p_\eps(\eps)$ with $p_\eps(\cdot)$ not dependent on $\l$. For example, if $\t\sim N(\mu,\Sigma)$, then $\t$ can be written as $\t=\mu+\Sigma^{1/2}\eps$ with $\eps\sim N(0,I)$.
Hence,
\[\LB(\l)=A(\l)+\E_{\eps\sim p_\eps}[\ell(g(\l,\eps))].\]
The gradient of the lower bound is 
\bean
\nabla_\l \LB(\l)&=&\nabla_\l A(\l)+\E_{\eps\sim p_\eps}\Big[\nabla_\l g(\l,\eps)\nabla_\t\ell(g(\l,\eps))\Big].
\eean
Note that if $A(\l)$ cannot be computed analytically or it is inconvenient to do so,
we can always represent the entire $\nabla_\l \LB(\l)$ as an expectation with respect to $\eps$.
By generating $\eps$ from $p_\eps(\cdot)$,
we are able to obtain an unbiased estimator $\widehat{\nabla_{\lambda}\LB}\left(\lambda\right)$
of the gradient $\nabla_{\lambda}\LB\left(\lambda\right)$.
Therefore, we can use stochastic optimization to optimize $\LB(\lambda)$.
The representation of the gradient $\nabla_\l \LB(\l)$ in terms of an expectation with respect to $\eps$ rather than $\theta$ is the so-called reparameterization trick \citep{Kingma:2013}.
In general it works more efficient than the alternative methods that sample directly from $q_\l(\t)$ \citep{Ruiz:2015,Tan:2017}.
One of the reasons is that this reparameterization takes into account the information from the gradient of the log-likelihood.

We now extend the method to the case where $\nabla_\t\ell(\t)$ is intractable but can be estimated unbiasedly.
Let $\wh G(\t)$ be an unbiased estimator of $\nabla_\t\ell(\t)$, i.e. $\E(\wh G(\t))=\nabla_\t\ell(\t)$,
where the expectation is with respect to all random variables $u$ needed for computing $\wh G(\t)=\wh G(\t,u)$.
Typically, $u$ is a set of uniform random numbers.
Write $p_U(\cdot)$ for the distribution of $u$. Then,
\bean
\nabla_\l \LB(\l)&=&\nabla_\l A(\l)+\E_{\eps\sim p_\eps,u\sim p_U}\Big[\nabla_\l g(\l,\eps)\wh G(\t,u)\Big], \;\;\t=g(\l,\eps),
\eean
which can be estimated unbiasedly by
\beq\label{eq:LB_gradient}
\wh{\nabla_\l \LB(\l)}=\nabla_\l A(\l)+\frac1S\sum_{i=1}^S\nabla_\l g(\l,\eps_i)\wh G(\t_i,u_i)
\eeq
where $\eps_i\sim p_\eps(\cdot)$, $u_i\sim p_U(\cdot)$, $\t_i=g(\l,\eps_i)$, $i=1,...,S$.
Therefore, we can use stochastic optimization \citep{Robbins:1951} to maximize $\LB(\lambda)$ as follows.
\begin{algorithm} \label{Algorithm1} 
\begin{itemize}
\item Set the number of samples $S$, initialize $\lambda^{(0)}$ and stop the following iteration if the
stopping criterion is met.
\item For $t=0,1,...$, compute $\lambda^{(t+1)}=\lambda^{(t)}+a_{t}I_F(\l^{(t)})^{-1}\widehat{\nabla_{\lambda}LB}\left(\lambda^{\left(t\right)}\right)$.
\end{itemize}
\end{algorithm}
Here $I_F(\l)=\cov_{q_\l}(\nabla_\l\log q_\l(\t))$ is the Fisher information matrix of $\l$ w.r.t. $q_\l(\t)$.
The sequence $\{a_{t},t\geq0\}$ is the learning rate and should satisfy
$a_{t}>0$, $\sum_{t}a_{t}=\infty$ and $\sum_{t}a_{t}^{2}<\infty$ \citep{Robbins:1951}.

In Algorithm \ref{Algorithm1} we use the natural gradient of the lower bound, which is defined as $I_F(\l)^{-1}{\nabla_{\lambda}\LB}\left(\lambda\right)$.
The natural gradient more 
adequately captures the geometry of the variational distribution $q_\lambda$; see, e.g., \cite{Amira:1998}.
Using this gradient often makes the convergence faster \citep{Tran:2015,Hoffman:2013}.
Here, we provide an informal explanation in the current context. 
It is easy to see that the Hessian matrix of the lower bound is
\[H(\lambda)=\nabla_{\lambda\lambda'}[\LB(\lambda)]=\int \nabla_{\lambda\lambda'}[\log q_\lambda (\theta)]q_\lambda(\theta)\log\frac{\pi(\theta)}{q_\lambda(\theta)}d\theta-I_F(\lambda),\]
which is approximately $-I_F(\lambda)$ when $q_\lambda\approx \pi$.
That is, Algorithm \ref{Algorithm1} is a Newton-Raphson type algorithm as it uses the second-order information of the target function.

It is important to note that VBILL is exact in the sense that it minimizes the same Kullback-Leibler divergence $\KL(q_\lambda\|\pi)$  
as the target that we would optimize if the exact likelihood is available.
Therefore, apart from an extra Monte Carlo error, the VBILL approach is able to produce estimators as if the true likelihood, or full-data likelihood, was used. 
The VBIL approach of \cite{Tran:2015} works on an augmented space and minimizes 
a Kullback-Leibler divergence that is equal to $\KL(q_\lambda\|\pi)$ only if the variance of the log of the estimated likelihood is constant.

Convergence properties of the stochastic optimization procedure in Algorithm \ref{Algorithm1} are well-known
in the literature \citep[see, e.g.,][]{Sacks:1958}.
Similarly to \cite{Tran:2015}, it is possible to show that the variance of VBILL estimators increases only linearly with the variance of the estimated gradient of the log-likelihood,
which suggests that VBILL is robust to variation in estimating this gradient.

\subsubsection*{Stopping rule}

As in \citet{Tran:2015}, the updating algorithm is stopped if the
change in the average value of the lower bounds over a window of $K$
iterations, 
\[\overline{\LB}\left(\lambda^{\left(t\right)}\right):=\frac{1}{K}\sum_{k=1}^{K}\widehat{\LB}\left(\lambda^{\left(t-k+1\right)}\right),\]
is less than some threshold $\varepsilon$, where $\widehat{\LB}\left(\lambda\right)$
is an estimate of $\LB\left(\lambda\right)$. Furthermore,
we also use the scaled version of the lower bound $\widehat{\LB}\left(\lambda\right)/n$,
with $n$ the size of the dataset. The scaled lower bound is roughly independent
of the size of the dataset. Our paper sets $K=5$. 
Alternatively, we can stop the updating algorithm if the
change in the average value $\overline{\lambda}^{\left(t\right)}=\left(1/K\right)\sum_{k=1}^{K}\lambda^{\left(t-k+1\right)}$,
is less than some threshold $\varepsilon$.
A byproduct of using the stopping rule based on the lower bound is that the lower bound estimates can be useful for model selection \citep{Sato:2001,Nott:2012}.

\subsection{VBILL with Data Subsampling}
This section presents the VBILL method for Big Data. 
Let $y=\{y_{i},i=1,...,n\}$ be the data set. We assume that the $y_i$ are independent so that the likelihood
is $L(\theta)=\prod\nolimits _{i=1}^{n}p(y_{i}|\theta)$ and we assume for now that each likelihood contribution $p(y_{i}|\theta)$
can be computed analytically.
The log-likelihood is 
\begin{equation}
\ell\left(\theta\right):=\sum_{i=1}^{n}\ell_{i}\left(\theta\right),\;\quad \text{where} \quad 
\ell_{i}\left(\theta\right):=\log p\left(y_{i}|\theta\right).\label{eq:log-likelihood formula}
\end{equation}
The gradient of the log-likelihood is
\[\nabla_\t\ell(\t)=\sum_i\nabla_\t\ell_i(\t),\]
where each $\nabla_\t\ell_i(\t)$ can be computed analytically or by using numerical differentiation.
We are concerned with the case where this gradient
is computationally intractable 
in the sense that $n$ is so large that computing this sum is impractical. 
The proposed VBILL approach to this problem
is based on the key observation that 
it is convenient and computationally much cheaper to obtain an unbiased estimator $\widehat{G}(\theta)$ of the gradient
of the log-likelihood $\nabla_{\theta}\ell\left(\theta\right)$.

Let $g_i(\t):=\nabla_\t\ell_i(\t)$.
Let $\overline{\theta}$ be some central value of $\theta$ obtained by using, for example, Maximum Likelihood or MCMC, based on a representative subset of the full data.
By the first-order Taylor series expansion of the vector field $g_i(\t)$ \citep[Chapter 8]{Apostol:1969},
\bea\label{eq:g_i_taylor}
g_i(\t)&=&g_i(\overline\t)+\nabla_{\t'} g_i(\overline\t)(\t-\overline\t)+o(\|\t-\overline\t\|)\notag\\
&=&\nabla_\t\ell_i(\overline\t)+\nabla_{\t\t'}^2\ell_i(\overline\t)(\t-\overline\t)+o(\|\t-\overline\t\|)=w_i(\theta)+o(\|\t-\overline\t\|),
\eea
where
\[w_i(\theta):=\nabla_\t\ell_i(\overline\t)+\nabla_{\t\t'}^2\ell_i(\overline\t)(\t-\overline\t),\]
and $o(\delta)$ denotes the small order of $\delta$, meaning $o(\delta)/\delta\to0$ as $\delta\to0$. We can write
\bean
\nabla_\t\ell(\t)&=&\sum_i w_i(\t)+\sum_i d_i(\t)=w(\t)+d(\t),
\eean
with $d_i(\t)=g_i(\t)-w_i(\t)$, $w(\theta)=\sum_i w_i(\t)$ and $d(\theta)=\sum_i d_i(\t)$. It is computationally cheap to compute the first term
\[w(\t)=\sum_i\nabla_\t\ell_i(\overline\t)+\Big(\sum_i\nabla_{\t\t'}^2\ell_i(\overline\t)\Big)(\t-\overline\t)=A(\overline\t)+B(\overline\t)(\t-\overline\t) \]
because the sums $A(\overline\t)$ and $B(\overline\t)$ are computed just once. The second term $d(\t)$ can be estimated unbiasedly by 
simple random sampling
\[
\widehat{d}_{m}(\t)=\frac{1}{m}\sum_{i=1}^{m}nd_{u_{i}}(\t),
\]
where $u=\left(u_{1},...,u_{m}\right)$, $u_{i}\in F=\{1,2,...,n\}$, is
the $m\times1$ vector of indices obtained by simple random
sampling with replacement from the full index set $F$, $\P(u_i=k)=1/n$ for all $k\in F$. 
Here $m<n$ is the size of subsamples.
It is easy to show that $\ensuremath{\E}\left(\widehat{d}_{m}(\t)\right)=d(\t)$. 
Therefore,
\begin{equation}
\widehat{G}\left(\theta,u\right):=w(\t)+\widehat{d}_{m}(\t)
\end{equation}
is an unbiased estimator of log-likelihood gradient $\nabla_\t\ell(\t)$. 

Since $w_{i}\left(\theta\right)$ is an approximation of $g_{i}\left(\theta\right)$,
the differences $d_{i}\left(\theta\right)$ should have roughly the same size across $i$. 
Thus, $\widehat{d}_{m}(\t)$ is expected to be an efficient estimator of $d(\theta)$ \citep{Quiroz:2015a}.
If $\overline{\t}$ is an MLE of $\theta$,
then by the Bernstein von Mises theorem \citep[see, e.g.][]{Chen:1985}, $\|\theta-\overline{\t}\|=O_P(n^{-1/2})$,
with $O_P$ the stochastically big order with respect to the posterior distribution of $\theta$.
Then, from \eqref{eq:g_i_taylor}, the $d_i(\t)$ are very small, and thus $\wh G(\t,u)$ has a small variance.
See also \cite{Bardenet2015} and \cite{Quiroz:2015a} who demonstrate the efficiency of data subsampling estimators.
This guarantees that the variance of the gradient \eqref{eq:LB_gradient} is small,
which makes the VBILL procedure highly efficient. 
   
\subsection{VBILL with Data Subsampling for Big Panel Data}\label{sec:VBILL-Big Panel}
This section describes the VBILL method for estimating 
Big Panel Data models. 
For panel data models with $n$ panels $\left\{ y_{1},...,y_{n}\right\} $, also called longitudinal models or generalized linear mixed models,
the log-likelihood is still in the form of \eqref{eq:log-likelihood formula}, but each likelihood contribution $p\left({y}_{i}|\theta\right)$ is an integral over random effects $\alpha_i$,
\begin{equation}
p\left({y}_{i}|\theta\right)=\int p\left({y}_{i}|\theta,{\alpha}_{i}\right)p\left({\alpha}_{i}|\theta\right)d{\alpha}_{i}.\label{eq:likelihood contribution}
\end{equation}
In many cases the integral \eqref{eq:likelihood contribution} is analytically intractable and
hence also its first and second derivatives.
Therefore, it is intractable to compute the gradients $\nabla_{\theta_i}\ell(\theta)$ and $\nabla_\theta\ell(\theta)$, even when $n$ is small.
However, it is still possible to estimate these gradients unbiasedly.

We first describe how to obtain an unbiased estimator
of the gradient of the log-likelihood contributions $\nabla_{\theta}l_{i}\left(\theta\right)$.
The gradient $\nabla_{\theta}l_{i}\left(\theta\right)$ can be written as
\bea\label{eq:score}
\nabla_{\theta}l_{i}(\theta)&=&\frac{1}{p(y_i|\theta)}\nabla_\theta \int p\left({y}_{i},{\alpha}_{i}|\theta\right)d{\alpha}_{i}\notag\\
&=&\frac{1}{p(y_i|\theta)} \int \nabla_\theta \log p\left({y}_{i},{\alpha}_{i}|\theta\right)\times p\left({y}_{i},{\alpha}_{i}|\theta\right)d{\alpha}_{i}\notag\\
&=&\int \nabla_\theta \log p\left({y}_{i},{\alpha}_{i}|\theta\right)\times p\left({\alpha}_{i}|{y}_{i},\theta\right)d{\alpha}_{i}.
\eea
The representation in \eqref{eq:score} is known in the literature as Fisher's identity \citep{Cappe:2005}.
Therefore, we can obtain an unbiased estimator of $\nabla_{\theta}l_{i}(\theta)$ by using importance sampling.
When $n$ is large, it is clear that we can use the data subsampling approach described in the previous section,
together with Fisher's identity in \eqref{eq:score}, to obtain an unbiased estimator of the gradient $\nabla_\theta\ell(\theta)$.

\subsection{Gaussian variational distribution with factor decomposition}
Our paper uses a $d$-variate Gaussian distribution $\N(\theta;\mu,\Sigma)$, with $d$ the number of parameters,
for the VB distribution $q_\l(\t)$. 
If necessary, all the parameters can be transformed so that 
it is appropriate to approximate the posterior of the transformed parameters by a normal distribution.
This simplifies the stochastic VB procedure and avoids the factorization assumption in conventional VB that ignores the posterior dependence between the parameter blocks.
Using a Gaussian approximation is motivated by the Bernstein von Mises theorem \citep{Chen:1985},
which states that the posterior of $\theta$ is approximately Gaussian
when $n$ is large.
Therefore, using a Gaussian variational distribution results in a highly accurate approximation of the posterior distribution.
 
We also use a factor decomposition for $\Sigma$,
\beq\label{eq:factor_decomposition}
\Sigma=BB'+c^2I_d,
\eeq
with $B$ a $d\times 1$-vector, $c$ a scalar and $I_d$ the $d\times d$ identity matrix. This factor decomposition helps to reduce the total number of VB parameters from $d+d(d+1)/2$ in the 
conventional parameterization $\lambda=(\mu,\Sigma)$ to $2d+1$ in the parameterization $\l=(\mu,B,c)$.
The factorization also helps overcome the problem of
obtaining a negative-definite $\Sigma$ if it is updated directly.
It is possible to use the factor decomposition \eqref{eq:factor_decomposition} where $B$ is a $d\times k$ matrix and $k$ is the number of factors, with $k$ chosen by some model selection criterion \citep{Tan:2017}.
We use $k=1$ in this paper.
Because $B$ is a vector, we are able to find a closed-form expression for the inverse of the Fisher information matrix $I_F(\l)$ used in Algorithm 1. This closed-form expression leads to a significant gain in computational efficiency in Big Data settings. The closed-form expression for $I_F(\l)^{-1}$ is in the Appendix.

\subsubsection*{Initializing $\l$}
Let $\wh\t_{n_\text{sub}}$ be an estimate of $\t$ based on a subsample of size $n_\text{sub}$ from the full data of size $n$,
and let $I_{F,n_\text{sub}}$ be the observed Fisher information matrix based on this subsample.
We estimate the observed Fisher information matrix based on the full data by $I_{F,n}:=(n/n_\text{sub})I_{F,n_\text{sub}}$.
Let $\wh\Sigma:=I_{F,n}^{-1}$.
Motivated by the Bernstein von Mises theorem \citep[see, e.g.][]{Chen:1985}, which states that the posterior of $\theta$ is approximately Gaussian when $n$ is large,
we initialize the VB distribution $q_\lambda(\theta)$ by the Gaussian distribution with mean $\wh\t_{n_\text{sub}}$
and covariance matrix $\wh\Sigma$.
Let $(\nu_i,v_i)$, $\nu_1\geq\nu_2\geq...$, be the pairs of eigenvalues $\nu_i$ with the corresponding eigenvectors $v_i$ of $\hat\Sigma$.
Because $\wh\Sigma=\sum_i\nu_iv_iv_i'$, we initialize $B$ by $B=\sqrt{\nu_1}v_1$ and $c$ by $c=[\sum(\diag(\wh\Sigma-BB'))/d]^{1/2}$. Then $BB'+c^2I_d\approx \hat\Sigma$. The mean $\mu$ is initialized by $\wh\t_{n_\text{sub}}$.

\subsection{Randomised Quasi Monte Carlo}
Numerical integration using quasi-Monte Carlo (QMC) methods has proved in many cases to be more efficient than standard Monte Carlo methods
\citep{Niederreiter:1992,Dick:2010,Glasserman:2004}.
Standard Monte Carlo estimates the integral of interest based on i.i.d points from the uniform distribution
${u}_{s}\sim U\left[0,1\right]^{d}$.
QMC chooses these points deterministically and more evenly in the sense that they minimize
the so-called star discrepancy of the point set. 
Randomized  QMC (RQMC) then  adds randomness  to  these  points  such  that  the  resulting
points preserve the low-discrepancy property and, at the same time, they have
a uniform distribution marginally.  
Introducing randomness into QMC points is important in order to obtain
statistical properties such as unbiasedness and central limit theorems.
Our paper generates RQMC numbers using the scrambled net method of \citet{Matousek:1998}.
We use RQMC to sample $\eps\sim p_{\eps}\left(\cdot\right)$.
For the panel data example, we also use RQMC to obtain unbiased estimates of the likelihood and the gradient of the log-likelihood. 

\section{Experimental studies}\label{sec:Experimental studies}
\subsection{The US airlines data}
We use the airline on-time performance data from the 2009 ASA Data Expo to demonstrate the VBILL methodology. This is a 
massive dataset that exceeds the memory (RAM) of a single desktop computer.
The dataset, used previously by \citet{C.Wang2015} and \citet{Kane2013} among others,
consists of the flight arrival and departure details for all commercial
flights within the USA, from October 1987 to April 2008. The full
dataset, ignoring the missing values, has 22,347,358 observations. 

The response variable of the logistic regression model is late arrival,
which is set to 1 if a flight is late by more than 15 minutes and
0 otherwise. There are three covariates. The two binary covariates
are: \texttt{night} (1 if the departure occurred at night and 0 otherwise)
and \texttt{weekend} (1 if the departure occurred on weekend and 0 otherwise).
The third covariate is \texttt{distance}, which is the distance from origin to destination (in 1000 miles). 

We first compare the performance of VBILL with data subsampling to MCMC for a subset of $1113638$ observations
from the full dataset. 
This ``moderate" data example allows us to run the ``gold standard" MCMC,
so that we are able to assess the accuracy of VBILL.
The MCMC chain, based on the adaptive random walk
Metropolis-Hastings algorithm in \citet{Haario:2001}, consists of
30000 iterates with another 10000 iterates used as burn-in. We also
compare the use of RQMC and MC. We use a diffuse normal prior $N\left(0,50I_{4}\right)$ for the coefficient vector $\beta$.

The central value $\overline{\theta}$ is set as the MLE of $\beta$ based on 30\% of the full dataset (the one with $1113638$ observations).
For the number of samples $S$ in \eqref{eq:LB_gradient},
we use $S=2^{8}$ for both MC and RQMC.
Choosing $S$ as a power of $2$ (with 2 the base of the Sobol' sequence in the scrambled net method of \citet{Matousek:1998})
is convenient for the RQMC method.
We run this example on a single desktop with 4 local processors. 
We set the threshold $\varepsilon=10^{-7}$.

\begin{figure}[H]
\caption{Logistic regression model ($n=1113638,\ m=10000$): Plots of the MCMC and VBILL estimates of the marginal posteriors $p(\beta_j|y)$.\label{fig:Logistic-Model-Estimation figure1}}

\centering{}\includegraphics[width=15cm,height=10cm]{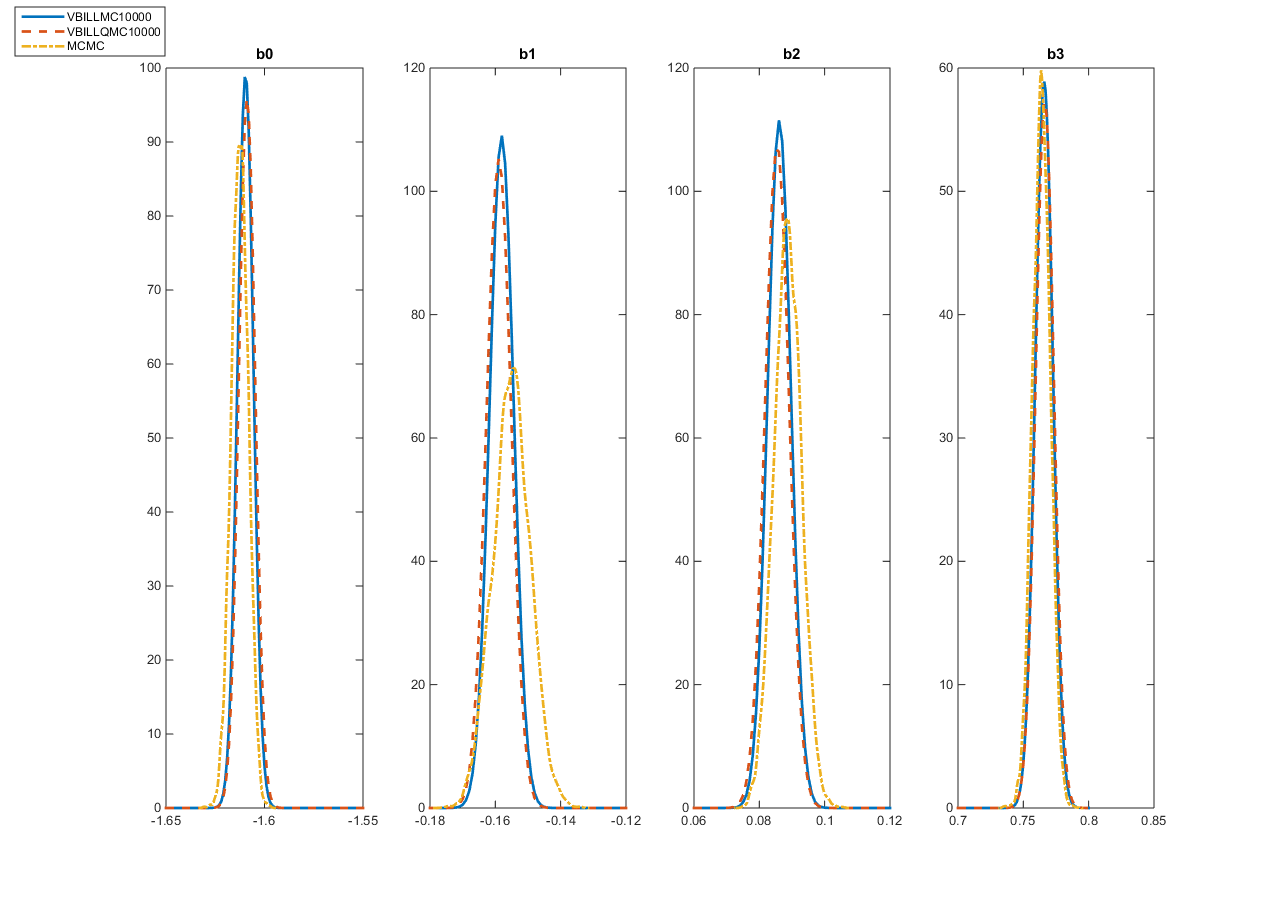}
\end{figure}

\begin{figure}[H]
\caption{Logistic regression model ($n=1113638,\ m=20000$): Plots of the MCMC and VBILL estimates of the marginal posteriors $p(\beta_j|y)$. \label{fig:Logistic-Model-Estimation fugure2}}

\centering{}\includegraphics[width=15cm,height=10cm]{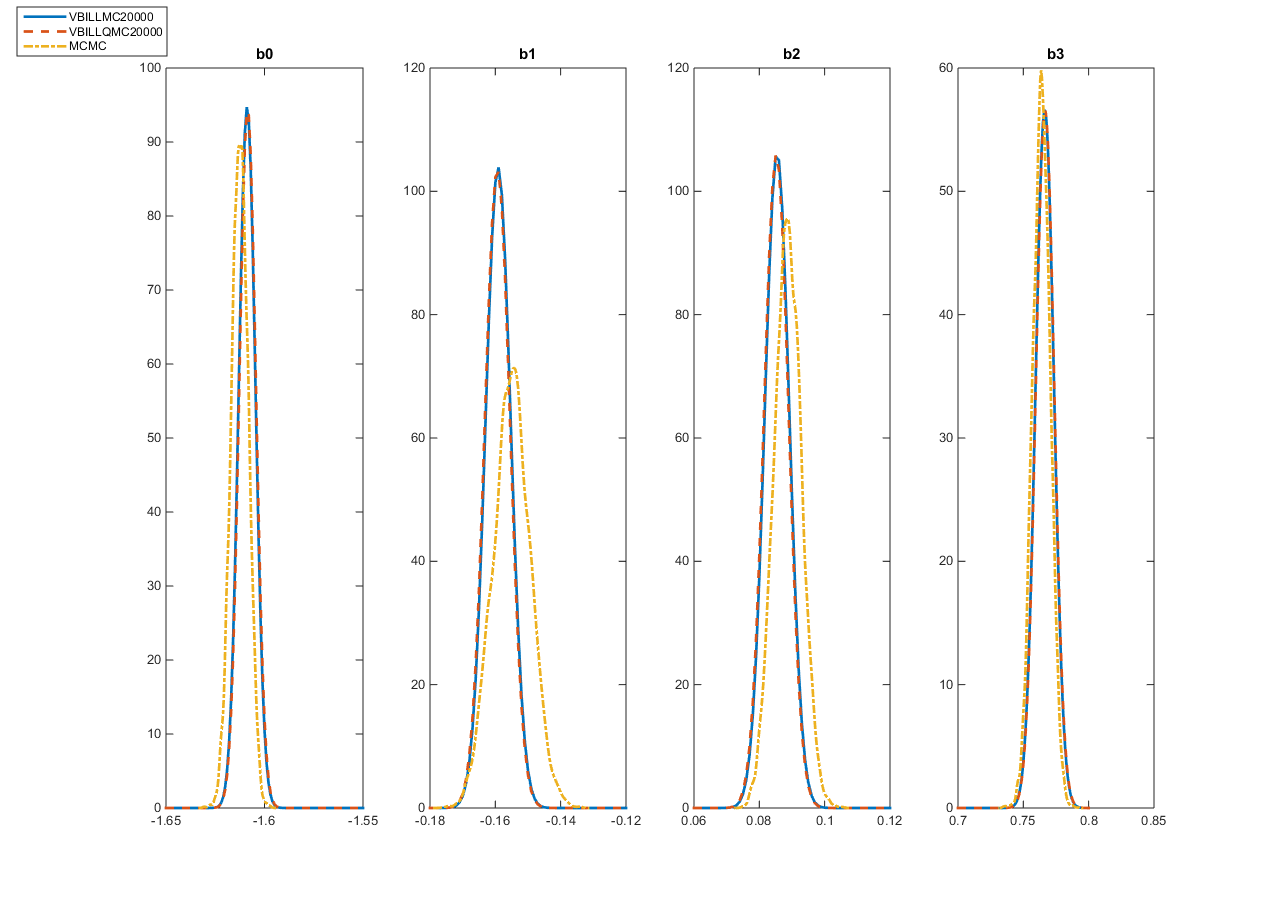}
\end{figure}

Table \ref{tab:Logistic-Model-Estimation sub 1million} summarizes the estimation results and, in particular, reports the
posterior means and posterior standard deviations of the parameters for
the VBILL and MCMC methods.
The results for VBILL
are obtained by averaging over 100 replications.
For VBILL, we use a subsample size $m=10000$ and $20000$, i.e. 1\% and 2\% of the full dataset. The table shows that the VBILL
estimates are very close to the ``gold standard'' MCMC estimates,
but that VBILL is around 20 times faster than MCMC in this moderate
data example. 
The more processors we have, the faster the VBILL
method will be. In general, VBILL with RQMC converges with fewer iterations than VBILL with MC. 
Figures \ref{fig:Logistic-Model-Estimation figure1}
and \ref{fig:Logistic-Model-Estimation fugure2} plot the MCMC and
VBILL estimates of the marginal posterior densities of $p\left(\beta|y\right)$.
The MCMC density estimates are obtained using the Matlab kernel density function \texttt{ksdensity}.
The figures show that the VBILL estimates are very close to the MCMC
estimates. 

To study the stability of VBILL,
Table \ref{table:std1} reports the standard errors, estimated over 100 replications, of the VBILL estimates of the posterior means and posterior standard deviations.
Clearly, the standard errors decrease when the subsample size $m$ increases. 
These standard errors suggest that the VBILL approach in this example is stable 
in the sense that the VBILL estimates across different runs stay the same up to at least the second decimal place.
We can also reduce these standard errors by using a smaller threshold $\varepsilon$.

\begin{table}[H]
\caption{Logistic regression model ($n=1113638$). The table summarizes the estimates of the posterior
means and the posterior standard deviations (in brackets). The VBILL
results are obtained by averaging over $100$ replications. The CPU time (in minutes) and the number of iterations for VBILL are averaged over the replications. 
\label{tab:Logistic-Model-Estimation sub 1million}}

\centering{}%
\begin{tabular}{cc|c|cc|cc}
 & &MCMC&\multicolumn{4}{c}{VBILL} \\
\hline
Parameter & $\overline{\theta}$ & & \multicolumn{2}{c|}{MC} &\multicolumn{2}{c}{RQMC}\tabularnewline
\hline 
$\beta_{0}$ & $-1.598$ & $\underset{\left(0.004\right)}{-1.613}$ & $\underset{\left(0.004\right)}{-1.609}$ & $\underset{\left(0.004\right)}{-1.609}$ & $\underset{\left(0.004\right)}{-1.609}$ & $\underset{\left(0.004\right)}{-1.609}$\tabularnewline
$\beta_{1}$ & $-0.175$ & $\underset{\left(0.006\right)}{-0.155}$ & $\underset{\left(0.004\right)}{-0.158}$ & $\underset{\left(0.004\right)}{-0.159}$ & $\underset{\left(0.004\right)}{-0.159}$ & $\underset{\left(0.004\right)}{-0.159}$\tabularnewline
$\beta_{2}$ & $0.051$ & $\underset{\left(0.004\right)}{0.089}$ & $\underset{\left(0.004\right)}{0.086}$ & $\underset{\left(0.004\right)}{0.085}$ & $\underset{\left(0.004\right)}{0.085}$ & $\underset{\left(0.004\right)}{0.085}$\tabularnewline
$\beta_{3}$ & $0.805$ & $\underset{\left(0.007\right)}{0.764}$ & $\underset{\left(0.007\right)}{0.766}$ & $\underset{\left(0.007\right)}{0.766}$ & $\underset{\left(0.007\right)}{0.766}$ & $\underset{\left(0.007\right)}{0.767}$\tabularnewline
\hline 
$m$ &  &  & 1\% & 2\%  & 1\% & 2\%\tabularnewline
\hline
Iter. &  & 40000 & 52 & 29 & 35 & 29\tabularnewline
\hline
CPU time &  & 20.23 & 1.52 & 1.03 & 1.07 & 0.99\tabularnewline
\end{tabular}
\end{table}

\begin{table}[H]
\caption{Logistic regression model ($n=1113638$). Monte Carlo standard
errors of the estimates over 100 replications. The results show that VBILL is stable 
in the sense that the VBILL estimates across different runs stay the same up to at least the second decimal place.
}\label{table:std1}
\centering{}%
\begin{tabular}{ccccc}
\hline 
Parameter & \multicolumn{2}{c}{VBILL-MC} & \multicolumn{2}{c}{VBILL-RQMC}\tabularnewline
\hline 
$\E(\beta_{0}|y)$ & $0.0037$ & $0.0027$ & $0.0035$ & $0.0029$\tabularnewline
$\E(\beta_{1}|y)$ & $0.0036$ & $0.0027$ & $0.0034$ & $0.0028$\tabularnewline
$\E(\beta_{2}|y)$ & $0.0036$ & $0.0025$ & $0.0033$ & $0.0027$\tabularnewline
$\E(\beta_{3}|y)$ & $0.0031$ & $0.0021$ & $0.0028$ & $0.0023$\tabularnewline
$\V(\beta_{0}|y)$ & $0.2808\times10^{-5}$ & $0.1552\times10^{-5}$ & $0.2103\times10^{-5}$ & $0.1639\times10^{-5}$\tabularnewline
$\V(\beta_{1}|y)$ & $0.2436\times10^{-5}$ & $0.1359\times10^{-5}$ & $0.1816\times10^{-5}$ & $0.1411\times10^{-5}$\tabularnewline
$\V(\beta_{2}|y)$ & $0.2356\times10^{-5}$ & $0.1321\times10^{-5}$ & $0.1771\times10^{-5}$ & $0.1392\times10^{-5}$\tabularnewline
$\V(\beta_{3}|y)$ & $0.6523\times10^{-5}$ & $0.3481\times10^{-5}$ & $0.4756\times10^{-5}$ & $0.3653\times10^{-5}$\tabularnewline
\hline 
$m$ & 1\% & 2\%  & 1\% &  2\%\tabularnewline
\hline 
\end{tabular}
\end{table}

We now run VBILL for the full dataset, which exceeds the memory of a
single desktop computer. 
Given our computational facilities, it is computationally infeasible to run MCMC to fit the model using such a large dataset. 
We use the MapReduce programming technique
in Matlab to overcome the computer's memory issue. MapReduce is available in the
R2014b release of Matlab. The MapReduce programming model has three components
\begin{itemize}
\item A \texttt{datastore} function that reads and organizes the dataset into small chunks for the ``map''
function.
\item A \texttt{map} function calculates the quantities of interest for
each individual chunk. MapReduce calls the map function
once for each data chunk organized by datastore.
\item A \texttt{reduce} function aggregates outputs from the map function
and produces the final results.
\end{itemize}
The \texttt{datastore} function splits the full dataset into $K$ chunks randomly, each fits
into the memory of a single desktop computer. The log-likelihood, its
gradient and Hessian can be decomposed as
\[
\ell\left(\theta\right)=\sum_{k=1}^{K}l^{\left(k\right)}\left(\theta\right),\;\;\;\;\nabla_{\theta}\ell\left(\theta\right)=\sum_{k=1}^{K}\nabla_{\theta}l^{\left(k\right)}\left(\theta\right),\;\;\;\;\nabla_{\theta\theta^{'}}^{2}\ell\left(\theta\right)=\sum_{k=1}^{K}\nabla_{\theta\theta^{'}}^{2}l^{\left(k\right)}\left(\theta\right),
\]
where $l^{\left(k\right)}\left(\theta\right)$, $\nabla_{\theta}l^{\left(k\right)}\left(\theta\right)$,
and $\nabla_{\theta\theta^{'}}^{2}l^{\left(k\right)}\left(\theta\right)$
are the log-likelihood contribution and its gradient and Hessian 
based on data chunk $k$. Similarly, we denote by $\widehat{l}_{m_{k}}\left(\theta\right)$
and $\widehat{\nabla_{\theta}l_{m_{k}}\left(\theta\right)}$ the unbiased
estimator of $l^{\left(k\right)}\left(\theta\right)$ and $\nabla_{\theta}l^{\left(k\right)}\left(\theta\right)$,
based on a random subset of size $m_{k}$ from data chunk $k$. 
The \texttt{map} function is used to calculate the
chunk based estimate $\widehat{l}_{m_{k}}\left(\theta\right)$ and
$\widehat{\nabla_{\theta}l_{m_{k}}\left(\theta\right)}$ for each
chunk $k$. Then, the
\texttt{reduce} function aggregates all the chunk-based unbiased estimates into the full data based
estimate of the gradient of the log-likelihood
\begin{equation}
\widehat{G}_{m}\left(\theta\right)=\sum_{k=1}^{K}\widehat{\nabla_{\theta}l_{m_{k}}\left(\theta\right)}.\label{eq:gradest}
\end{equation}
Then, 
\[
\E\left(\widehat{G}_{m}\left(\theta\right)\right)=\sum_{k=1}^{K}\E\left(\widehat{\nabla_{\theta}l_{m_{k}}\left(\theta\right)}\right)=\sum_{k=1}^{K}\nabla_{\theta}l^{\left(k\right)}\left(\theta\right)=\nabla_{\theta}l\left(\theta\right).
\]
We note that this method is computer-memory efficient in the sense that
the full dataset does not need to remain on-hold and can be stored in different places.
It is important to note that our VBILL estimator is mathematically justified and independent
of data chunking, as the estimator $\widehat{G}_{m}\left(\theta\right)$ in \eqref{eq:gradest} is guaranteed to be unbiased. 

\begin{figure}[H]
\caption{Logistic regression model ($n=22,347,358$).\label{fig:Logistic-Model-Estimation full sample}}

\centering{}\includegraphics[width=15cm,height=10cm]{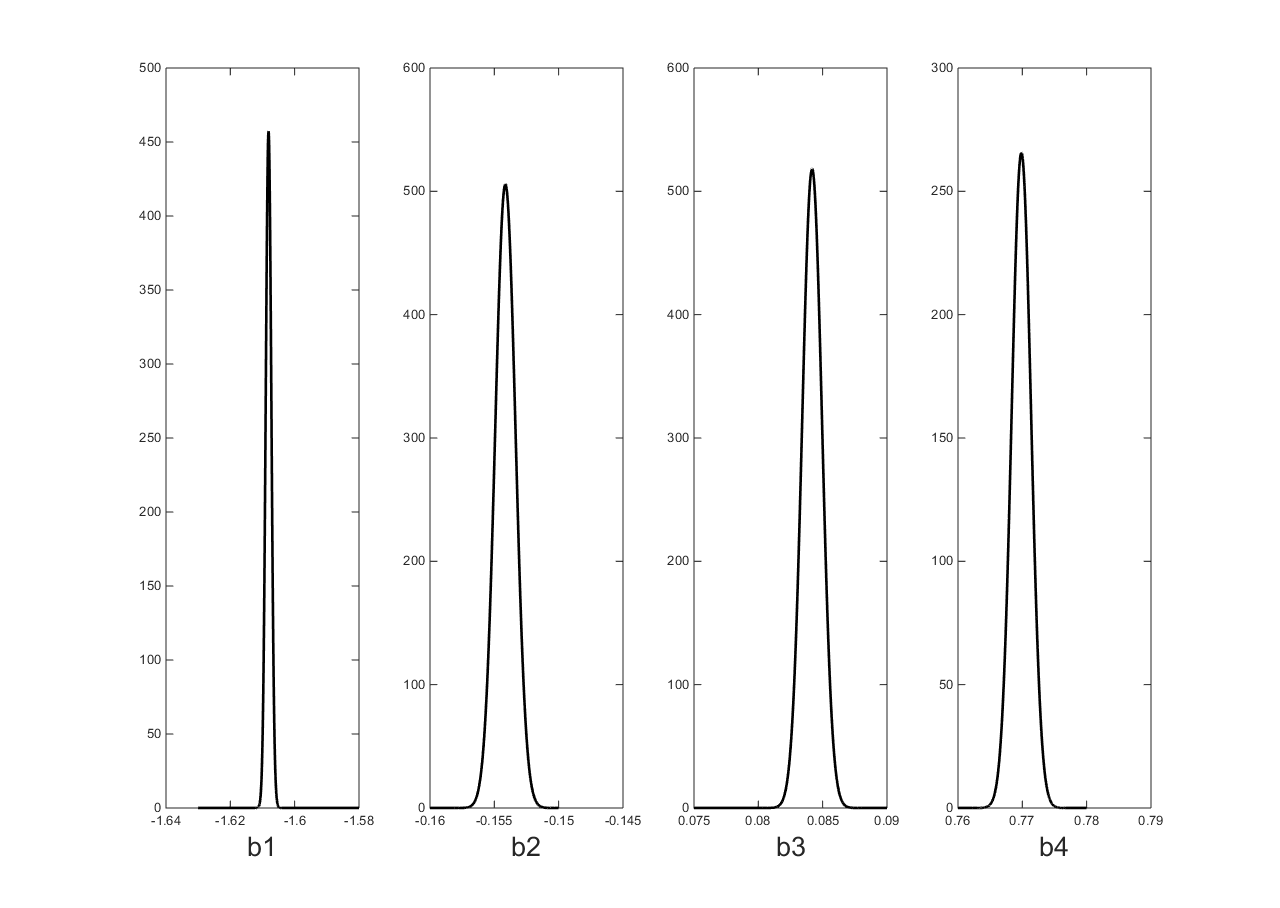}
\end{figure}

The central value $\overline{\theta}$ is the MLE based on 1 million observations of the full
dataset and is given in Table \ref{tab:Logistic-Model-Estimation full sample}.
We use approximately 5\% of the data in each subset. 
We estimate the gradient of the lower bound using RQMC and set $S=2^{8}$.
The VBILL method stopped after 24 iterations and the running time was 77.55 minutes.
Table \ref{tab:Logistic-Model-Estimation full sample}
shows the results and Figure \ref{fig:Logistic-Model-Estimation full sample} shows the
marginal posterior density estimates of the parameters, which are
bell-shaped with a very small variance as expected with a large 
dataset. 
This example demonstrates that the VBILL methodology is useful
for Bayesian inference for Big Data.

\begin{table}[H]
\caption{Logistic regression model ($n=22,347,358$).
Estimates of the posterior means and the posterior standard deviations (in brackets). 
The CPU time is in minutes.\label{tab:Logistic-Model-Estimation full sample}}

\centering{}%
\begin{tabular}{ccc}
\hline 
Param. & $\overline{\theta}$ & VBILL-RQMC\tabularnewline
\hline 
$\beta_{0}$ & $-1.613$ & $\underset{\left(0.0008\right)}{-1.608}$\tabularnewline
$\beta_{1}$ & $-0.155$ & $\underset{\left(0.0008\right)}{-0.154}$\tabularnewline
$\beta_{2}$ & $0.088$ & $\underset{\left(0.0008\right)}{0.084}$\tabularnewline
$\beta_{3}$ & $0.764$ & $\underset{\left(0.0015\right)}{0.770}$\tabularnewline
\hline 
Iter. &  & 24\tabularnewline
CPU time &  & 77.55\tabularnewline
\hline 
\end{tabular}
\end{table}

\subsection{Simulation Study: Panel Data Model }
This section studies the performance of the VBILL method for the panel data model.
Data are generated from the following logistic model with a random intercept
\bean
p\left(y_{it}|\beta,\alpha_{i}\right)&=&\text{Binomial}\left(1,p_{it}\right),\\
\text{logit}\left(p_{it}\right)&=&x_{it}'\beta+\alpha_{i},\;\;\alpha_{i}\sim N\left(0,\tau^{2}\right),
\eean
for $i=1,...,n$ and $t=1,...,5$.
We generate two datasets of $n=1000$ and $n=10000$ with $x_{1,it},...,x_{10,it}\sim U\left(0,1\right)$.
Let $\gamma=\log\left(\tau^{2}\right)$, so the model parameters are
$\theta=\left(\beta,\gamma\right)$. 
We use a diffuse normal prior $N\left(0,50I_{d}\right)$ for $\theta$; $d=12$ in this example.

The performance of VBILL is compared to the pseudo-marginal MCMC simulation method \citep{Andrieu:2009},
which is still able to generate samples from the posterior when the likelihood
in the Metropolis-Hastings algorithm is replaced by its unbiased estimator.
The likelihood in the panel data context is a product of $n$ integrals
over the random effects. Each integral is estimated unbiasedly using importance
sampling, with the number of importance samples chosen such that
the variance of unbiased likelihood estimator is approximately 1 \citep{Pitt:2012}. Each MCMC
chain consists of 30000 iterates with another 10000 used
as burn-in iterates. 

\begin{figure}[H]
\caption{Panel data example ($n=1000$): comparing MCMC and VBILL estimates
with $m=50$. \label{fig:Simulation-Results-for logistic random effect 50}}

\centering{}\includegraphics[width=15cm,height=10cm]{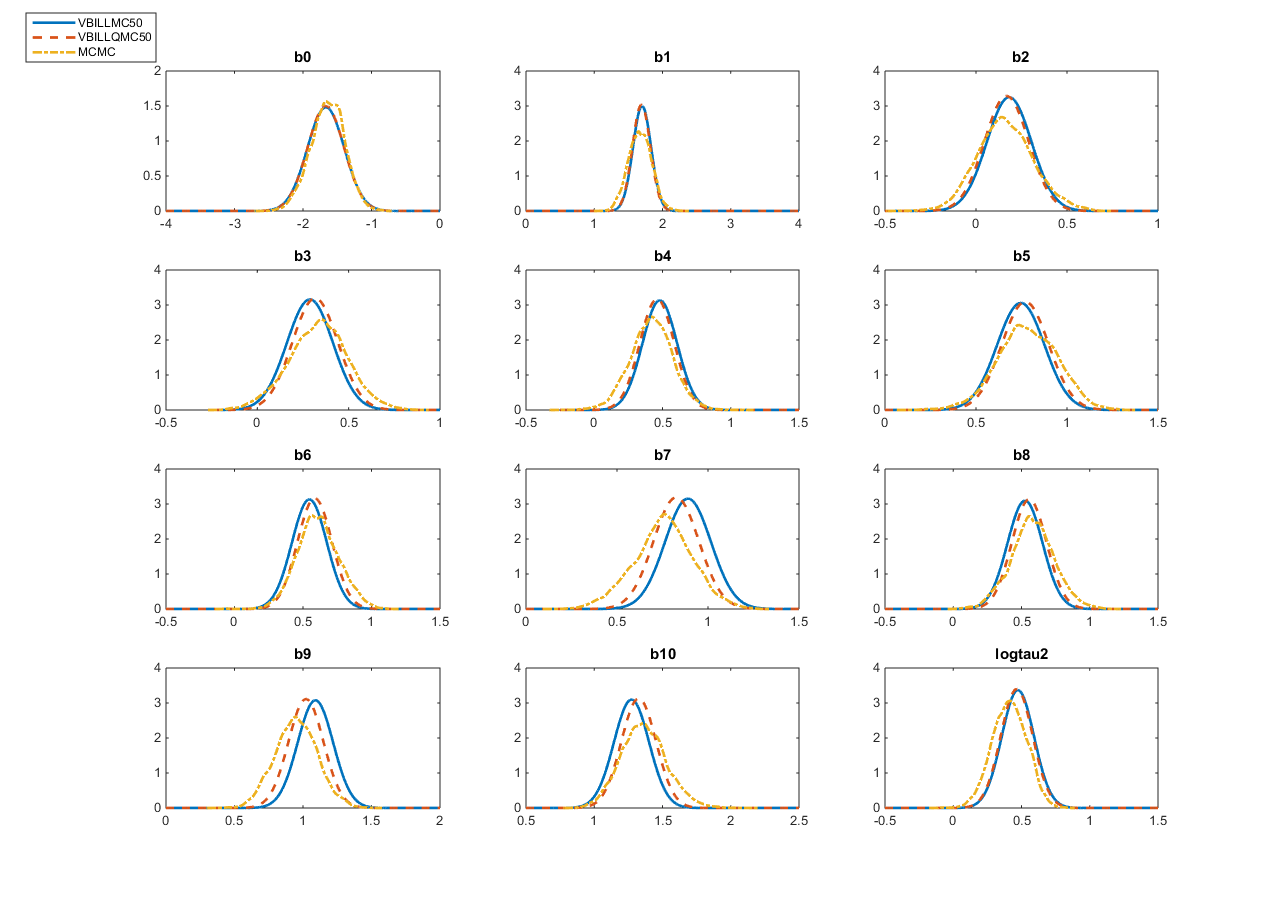}
\end{figure}

\begin{figure}[H]
\caption{Panel data example ($n=1000$): comparing MCMC and VBILL estimates
with $m=100$. \label{fig:Simulation-Results-for logistic random effect 100}}

\centering{}\includegraphics[width=15cm,height=10cm]{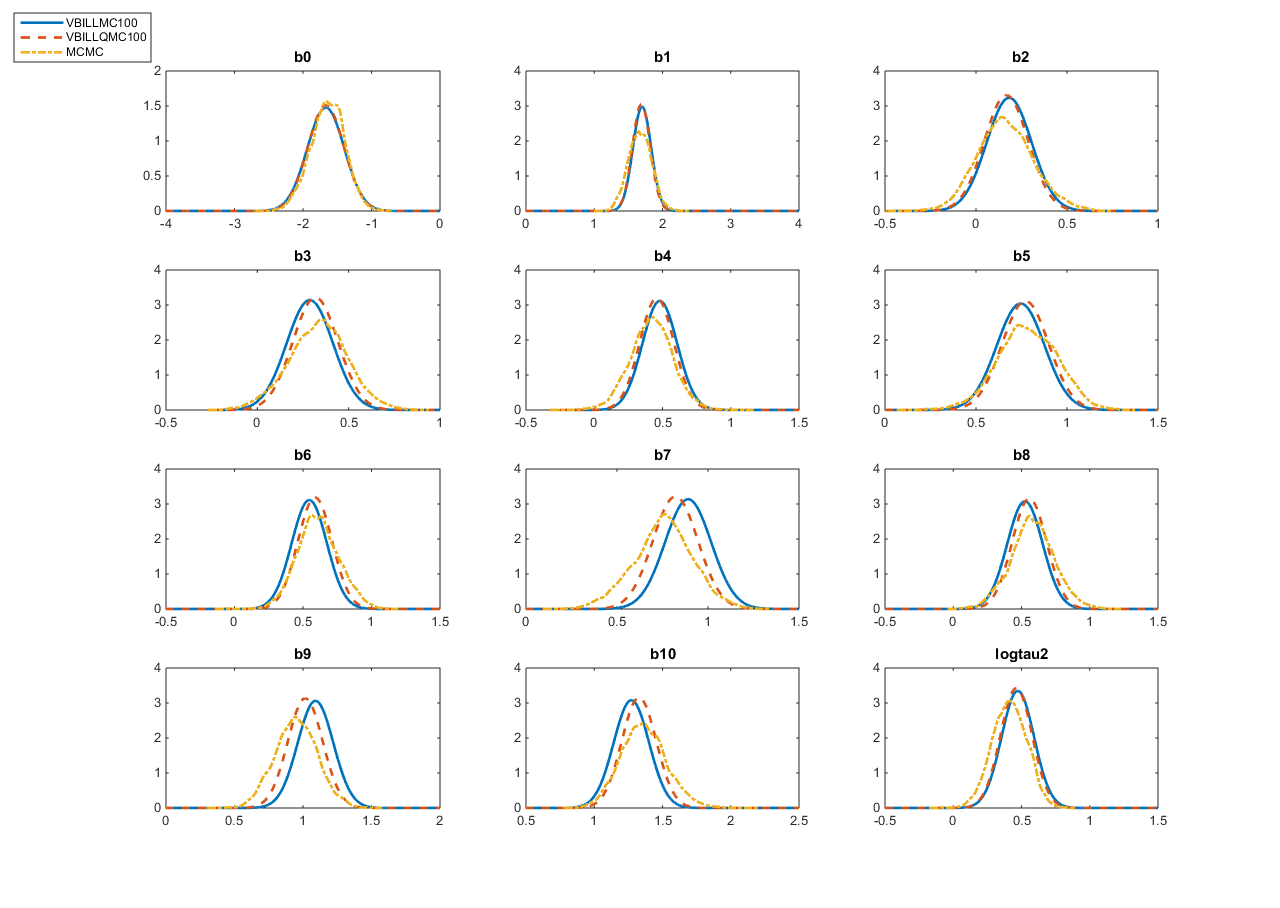}
\end{figure}

\begin{figure}[H]
\caption{Panel data example ($n=1000$): comparing the MCMC and VBILL estimates
with $m=200$. \label{fig:Simulation-Results-for logistic random effect 200}}

\centering{}\includegraphics[width=15cm,height=10cm]{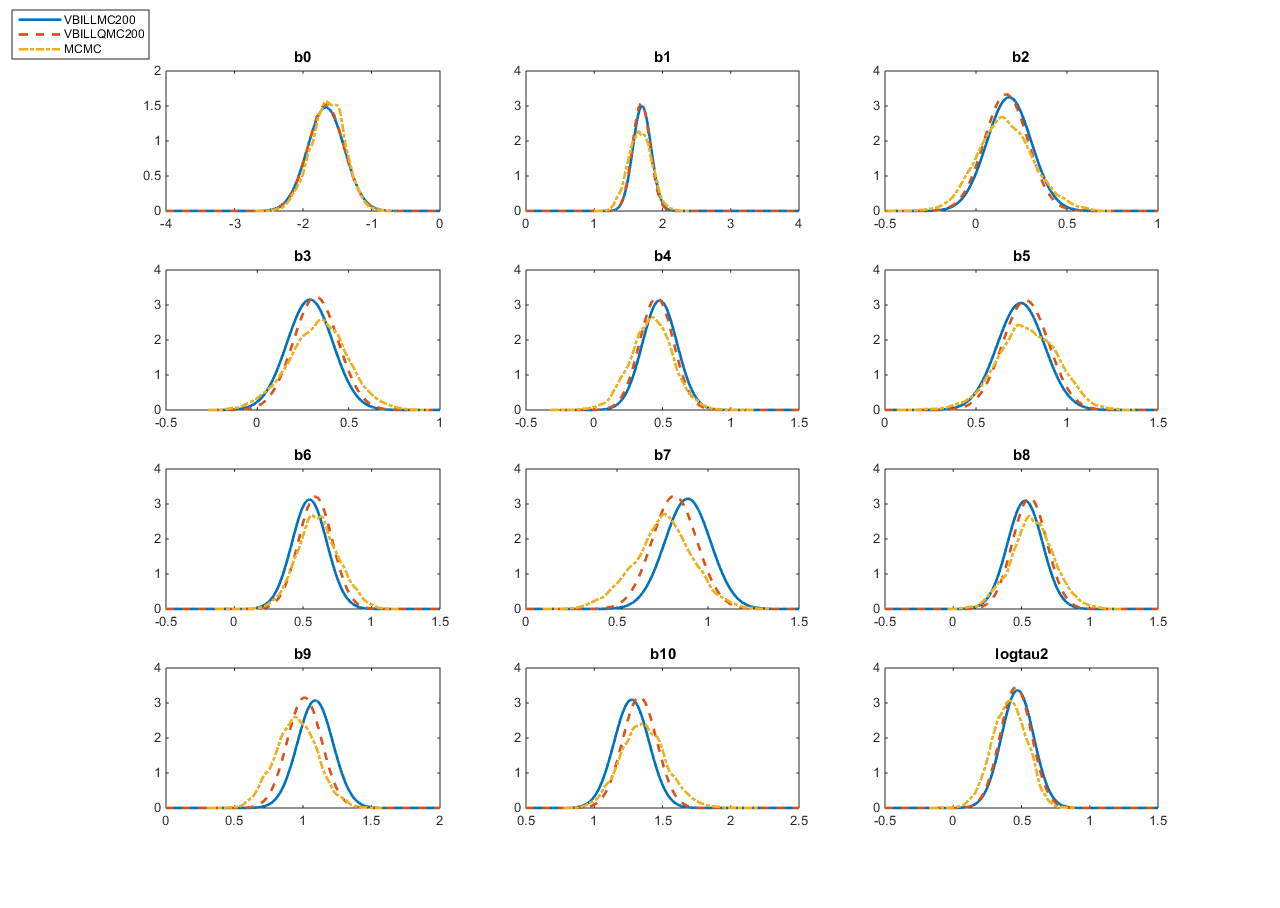}
\end{figure}

For VBILL, the central value $\overline{\theta}$
is a simulated maximum likelihood estimate of $\theta$ based on a 30\% randomly selected
subset of the full dataset. 
We set $S=2^{8}$ and use both MC and RQMC in VBILL. 
The number of importance samples used to estimate integrals in \eqref{eq:score} is $N=2^8$,
and $\varepsilon=10^{-5}$. 
Tables \ref{tab:Summary-of-Simulation logistic random effect 1000}
summarizes the performance results for the three methodologies:
pseudo-marginal MCMC, VBILL-MC and VBILL-RQMC
for various values of subsample size $m$. For VBILL, the results are obtained
by averaging over 100 replications. The VBILL estimates are close
to the MCMC estimates and they are all close to the true values.
However, VBILL is much faster than MCMC.
Figures \ref{fig:Simulation-Results-for logistic random effect 50},
\ref{fig:Simulation-Results-for logistic random effect 100}, and
\ref{fig:Simulation-Results-for logistic random effect 200} plot
the VBILL and MCMC estimates of the marginal posterior of the parameters.
The three figures show that the VBILL marginal posterior estimates
are very close to the MCMC estimates for both MC and RQMC and for
all subsample sizes. 
We note that VBILL with RQMC takes longer to run
compared to VBILL with MC, with not much difference in the number
of iterations and also the resulting marginal posterior
estimates in this example. This is because generating RQMC numbers takes
a longer time than generating plain pseudo random numbers. 

\begin{sidewaystable}
\caption{Panel data example ($n=1000$): Posterior mean
estimates (with posterior standard deviation in brackets). The VBILL
results are obtained by averaging over $100$ replications. The CPU time (in minutes) and the number of iterations for VBILL are averaged over 100 replications. 
\label{tab:Summary-of-Simulation logistic random effect 1000}}

\begin{centering}
\begin{tabular}{ccc|c|ccc|ccc}
 & & & MCMC&\multicolumn{6}{c}{VBILL} \\
\hline 
Param. & True & $\overline{\theta}$ &  & \multicolumn{3}{c|}{MC} &\multicolumn{3}{c}{RQMC}\\
\hline 
$\beta_{0}$ & $-1.5$ & $-1.72$ & $\underset{\left(0.25\right)}{-1.64}$ & $\underset{\left(0.27\right)}{-1.67}$ & $\underset{\left(0.27\right)}{-1.67}$ & $\underset{\left(0.27\right)}{-1.67}$ & $\underset{\left(0.27\right)}{-1.670}$ & $\underset{\left(0.26\right)}{-1.67}$ & $\underset{\left(0.26\right)}{-1.67}$\tabularnewline
$\beta_{1}$ & $1.5$ & $1.73$ & $\underset{\left(0.17\right)}{1.67}$ & $\underset{\left(0.13\right)}{1.70}$ & $\underset{\left(0.14\right)}{1.70}$ & $\underset{\left(0.13\right)}{1.70}$ & $\underset{\left(0.13\right)}{1.69}$ & $\underset{\left(0.13\right)}{1.69}$ & $\underset{\left(0.13\right)}{1.69}$\tabularnewline
$\beta_{2}$ & $0.5$ & $0.22$ & $\underset{\left(0.15\right)}{0.16}$ & $\underset{\left(0.12\right)}{0.18}$ & $\underset{\left(0.12\right)}{0.18}$ & $\underset{\left(0.12\right)}{0.18}$ & $\underset{\left(0.12\right)}{0.17}$ & $\underset{\left(0.12\right)}{0.17}$ & $\underset{\left(0.12\right)}{0.17}$\tabularnewline
$\beta_{3}$ & $0.25$ & $0.22$ & $\underset{\left(0.16\right)}{0.33}$ & $\underset{\left(0.13\right)}{0.29}$ & $\underset{\left(0.12\right)}{0.29}$ & $\underset{\left(0.12\right)}{0.29}$ & $\underset{\left(0.13\right)}{0.32}$ & $\underset{\left(0.12\right)}{0.32}$ & $\underset{\left(0.12\right)}{0.32}$\tabularnewline
$\beta_{4}$ & $0.3$ & $0.52$ & $\underset{\left(0.16\right)}{0.42}$ & $\underset{\left(0.13\right)}{0.48}$ & $\underset{\left(0.13\right)}{0.48}$ & $\underset{\left(0.13\right)}{0.48}$ & $\underset{\left(0.13\right)}{0.46}$ & $\underset{\left(0.12\right)}{0.46}$ & $\underset{\left(0.12\right)}{0.46}$\tabularnewline
$\beta_{5}$ & $0.8$ & $0.69$ & $\underset{\left(0.16\right)}{0.78}$ & $\underset{\left(0.13\right)}{0.75}$ & $\underset{\left(0.13\right)}{0.75}$ & $\underset{\left(0.13\right)}{0.75}$ & $\underset{\left(0.13\right)}{0.77}$ & $\underset{\left(0.13\right)}{0.77}$ & $\underset{\left(0.13\right)}{0.77}$\tabularnewline
$\beta_{6}$ & $0.45$ & $0.46$ & $\underset{\left(0.15\right)}{0.61}$ & $\underset{\left(0.13\right)}{0.55}$ & $\underset{\left(0.13\right)}{0.55}$ & $\underset{\left(0.13\right)}{0.55}$ & $\underset{\left(0.13\right)}{0.59}$ & $\underset{\left(0.13\right)}{0.59}$ & $\underset{\left(0.12\right)}{0.59}$\tabularnewline
$\beta_{7}$ & $0.85$ & $1.03$ & $\underset{\left(0.16\right)}{0.75}$ & $\underset{\left(0.13\right)}{0.89}$ & $\underset{\left(0.13\right)}{0.89}$ & $\underset{\left(0.13\right)}{0.89}$ & $\underset{\left(0.13\right)}{0.83}$ & $\underset{\left(0.12\right)}{0.82}$ & $\underset{\left(0.12\right)}{0.82}$\tabularnewline
$\beta_{8}$ & $0.75$ & $0.46$ & $\underset{\left(0.16\right)}{0.58}$ & $\underset{\left(0.13\right)}{0.53}$ & $\underset{\left(0.13\right)}{0.53}$ & $\underset{\left(0.13\right)}{0.53}$ & $\underset{\left(0.13\right)}{0.56}$ & $\underset{\left(0.13\right)}{0.56}$ & $\underset{\left(0.13\right)}{0.56}$\tabularnewline
$\beta_{9}$ & $0.67$ & $1.24$ & $\underset{\left(0.15\right)}{0.95}$ & $\underset{\left(0.13\right)}{1.09}$ & $\underset{\left(0.13\right)}{1.09}$ & $\underset{\left(0.13\right)}{1.09}$ & $\underset{\left(0.13\right)}{1.02}$ & $\underset{\left(0.13\right)}{1.02}$ & $\underset{\left(0.13\right)}{1.01}$\tabularnewline
$\beta_{10}$ & $1.5$ & $1.15$ & $\underset{\left(0.17\right)}{1.36}$ & $\underset{\left(0.13\right)}{1.27}$ & $\underset{\left(0.13\right)}{1.27}$ & $\underset{\left(0.13\right)}{1.27}$ & $\underset{\left(0.13\right)}{1.33}$ & $\underset{\left(0.13\right)}{1.33}$ & $\underset{\left(0.13\right)}{1.33}$\tabularnewline
$\gamma$ & $0.41$ & $0.60$ & $\underset{\left(0.13\right)}{0.41}$ & $\underset{\left(0.12\right)}{0.47}$ & $\underset{\left(0.12\right)}{0.47}$ & $\underset{\left(0.12\right)}{0.47}$ & $\underset{\left(0.12\right)}{0.47}$ & $\underset{\left(0.12\right)}{0.46}$ & $\underset{\left(0.12\right)}{0.46}$\tabularnewline
\hline 
$m$ &  &  &  & 50 & 100 & 200 &  50 & 100 & 200\tabularnewline
Iter. &  &  & 40000 & 24 & 22 & 22 & 25 & 26 & 26\tabularnewline
CPU time &  &  & 80 & 1.09 & 1.62 & 2.86 & 1.88 & 3.12 & 5.40\tabularnewline
\hline 
\end{tabular}
\par\end{centering}

\end{sidewaystable}

\subsubsection*{Larger Panel Data Example}
This section describes a scenario where it is difficult to use the
pseudo-marginal MCMC method. We consider a large data case with the number of panels
$n=10000$. The MCMC method is extremely expensive since the variance of the unbiased
estimator of the likelihood is large and it requires approximately
$N=8000$ importance samples in order to target the optimal variance
of 1 \citep{Pitt:2012}. Hence, if an optimal MCMC procedure is run on our computer to generate
40000 iterations, it would take $4053.30$ minutes. We run the VBILL-MC
and VBILL-RQMC with the subsample size $m=500$ and $\varepsilon=10^{-6}$. Table \ref{tab:Summary-of-Simulation logistic random effect10000}
summarizes the results as averages over 100 replications.
Both methods converge on average after 28
and 25 iterations respectively, and result in similar estimates of the posterior mean and standard deviation.
The times taken are 8.65 and 13.14 minutes for VBILL-MC
and VBILL-RQMC, respectively. 
The MC errors of the estimates, based on 100 replications,
suggest that using RQMC to estimate integrals in \eqref{eq:score}
helps to stablize the VBILL estimates.

\begin{figure}[H]
\caption{Panel data example ($n=10000$): comparing the MCMC and VBILL
estimates with $m=500$\label{fig:Simulation-Results-for logistic panel data 10000}}

\centering{}\includegraphics[width=15cm,height=10cm]{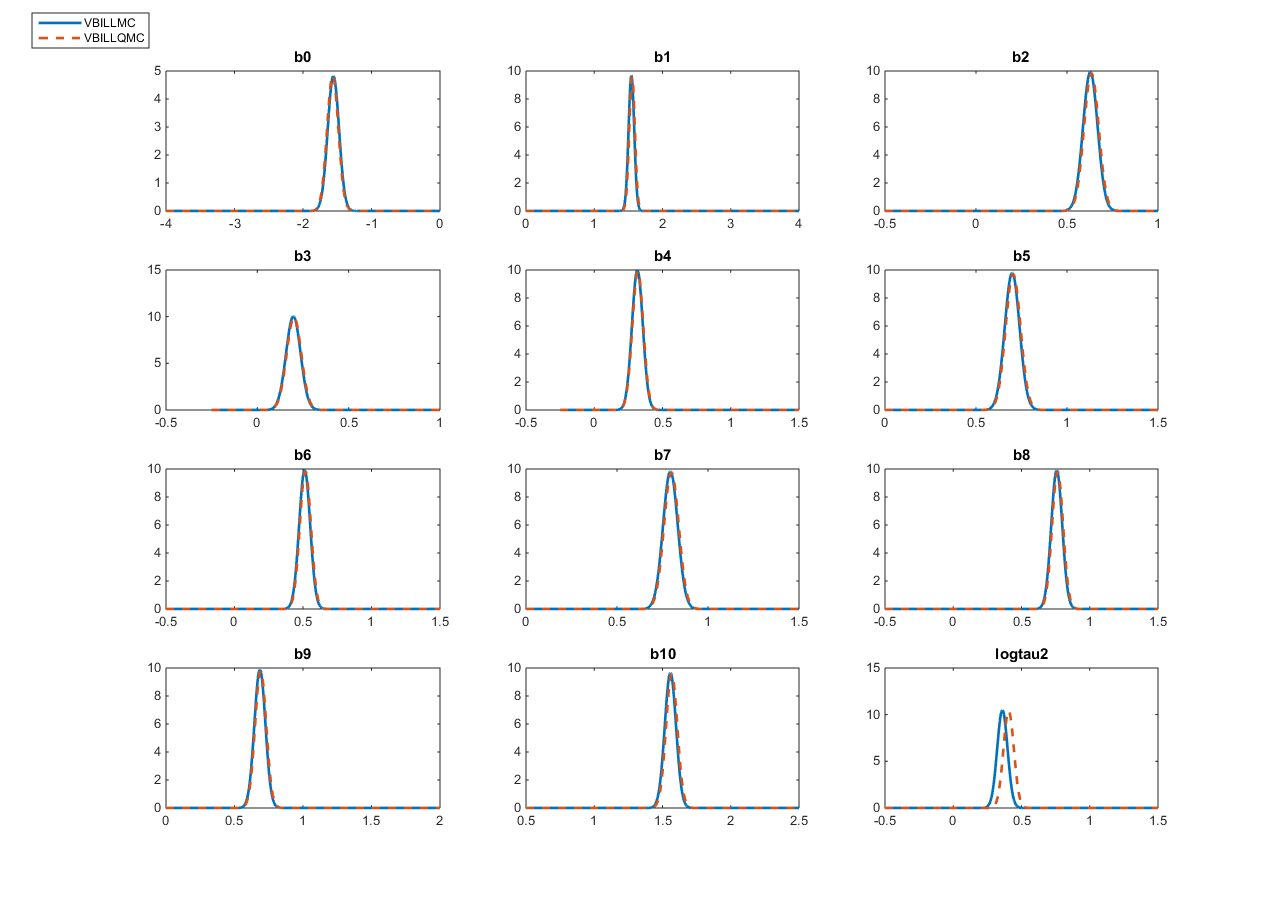}
\end{figure}

Figure \ref{fig:Simulation-Results-for logistic panel data 10000}
plots the variational approximations of the marginal posteriors of
the parameters, which are bell-shaped with small posterior variance
as expected with a very large dataset. The two methods produce very
similar results. This example demonstrates that VBILL offers a computationally efficient approach for Big Panel Data problems. 

\begin{table}[H]
\caption{Panel data example ($n=10000$): The table shows the estimates of posterior mean (first line) and posterior standard deviation (second line). The numbers in brackets are MC standard errors over $100$ replications, which suggests that VBILL-RQMC is more stable than VBILL-MC. 
The CPU time and number of iterations are averaged over the replications. 
\label{tab:Summary-of-Simulation logistic random effect10000}}
\centering{}%
\begin{tabular}{cccll}
\hline 
Param. & True & $\overline{\theta}$ & VBILL-MC & VBILL-RQMC\tabularnewline
\hline 
$\beta_{0}$ & $-1.5$ & $-1.71$ & -1.56 (0.0261)	 &-1.57 (0.0178)\tabularnewline
	    &        &         & 0.083 ($0.82\times10^{-3}$)   &  0.083 ($0.82\times10^{-3}$) \tabularnewline
$\beta_{1}$ & $1.5$  & $1.60$  & 1.54 ($0.0107$)	 &1.55 ($0.0064$)\tabularnewline
	    &  	     &         & 0.041 ($0.30\times10^{-3}$) &0.042 ($0.25\times10^{-3}$)\tabularnewline
$\beta_{2}$ & $0.5$ & $0.71$ & ${0.63} (0.0171)$ & ${0.63}$ ($0.0133$)\tabularnewline
&&& 0.040 ($0.29\times10^{-3}$)&0.041 ($0.25\times10^{-3}$)\tabularnewline
$\beta_{3}$ & $0.25$ & $0.24$ & ${0.20}$ (0.0069) & ${0.20}$ (0.0057)\tabularnewline
&&& 0.040 ($0.28\times10^{-3}$)&0.040 ($0.24\times10^{-3}$)\tabularnewline
$\beta_{4}$ & $0.3$ & $0.34$ & ${0.32}$ (0.0039) & ${0.32}$ (0.0030)\tabularnewline
&&&0.040 ($0.29\times10^{-3}$)&0.040 ($0.24\times10^{-3}$)\tabularnewline
$\beta_{5}$ & $0.8$ & $0.69$ & ${0.70}$ (0.0041) & ${0.70}$ (0.0033)\tabularnewline
&&&0.041 ($0.29\times10^{-3}$)&0.041 ($0.25\times10^{-3}$)\tabularnewline
$\beta_{6}$ & $0.45$ & $0.56$ & ${0.51}$ (0.0087) & ${0.51}$ (0.0066)\tabularnewline
&&&0.040 ($0.29\times10^{-3}$)&0.041 ($0.25\times10^{-3}$)\tabularnewline
$\beta_{7}$ & $0.85$ & $0.80$ & ${0.79}$ (0.0022) & ${0.80}$ (0.0008)\tabularnewline
&&&0.041 ($0.29\times10^{-3}$)&0.041 ($0.25\times10^{-3}$)\tabularnewline
$\beta_{8}$ & $0.75$ & $0.79$ & ${0.76}$ (0.0059) & ${0.76}$ (0.0037)\tabularnewline
&&&0.040 ($0.29\times10^{-3}$)&0.041 ($0.25\times10^{-3}$)\tabularnewline
$\beta_{9}$ & $0.67$ & $0.70$ & ${0.69}$ (0.0021) & ${0.69}$ (0.0006)\tabularnewline
&&&0.040 ($0.29\times10^{-3}$)&0.041 ($0.25\times10^{-3}$)\tabularnewline
$\beta_{10}$ & $1.5$ & $1.56$ & ${1.56}$ (0.0028) & ${1.57}$ (0.0034)\tabularnewline
&&&0.041 ($0.30\times10^{-3}$)&0.042 ($0.25\times10^{-3}$)\tabularnewline
$\gamma$ & $0.41$ & $0.35$ & $0.36$ (0.0117) & $0.41$ (0.0070)\tabularnewline
&&&0.038 ($0.27\times10^{-3}$)&0.038 ($0.23\times10^{-3}$)\tabularnewline
\hline 
$m$ &  &  & 500 & 500\tabularnewline
Iteration &  &  & 28 & 25\tabularnewline
CPU time (min) &  &  & 8.65 & 13.14\tabularnewline
\hline 
\end{tabular}
\end{table}

\section{Conclusions}\label{sec:Conclusions}
We propose the VBILL approach for Bayesian inference
when the likelihood function is intractable but the gradient of the log-likelihood can be estimated unbiasedly.
The method is useful for Big Data situations where it is convenient to obtain unbiased estimates 
of the gradient of the log-likelihood.
 
Unlike MCMC approaches that can in principle sample from the exact posterior,
VBILL, as a variant of VB, is an approximate method for estimating the posterior distribution of the parameters. 
The main advantage of VBILL is that it is much more computationally efficient than MCMC, but produces very similar results
to MCMC as shown in the simulated and real examples. 
To the best of our knowledge, in Big Data situations, there is no MCMC approach in the current literature 
that is both computationally efficient and able to sample from the exact posterior.
Conventional MCMC is exact, but is well-known to be extremely expensive in Big Data,
while data subsampling MCMC can be computationally efficient but is not exact \citep{Bardenet2015,Quiroz:2015a}.
For these reasons, we believe that VBILL can be the method of choice for Big Data applications. 

\section*{Acknowledgement}
The work of David Gunawan and Robert Kohn was partially supported by
an ARC Center of Excellence Grant CE140100049 and an ARC Discovery
grant DP150104630. We would like to thank two anonymous referees and
an annonymous associate editor for greatly helping to improve the content
and presentation of the paper.

\section*{Appendix}
\subsection*{Closed-form expression for $I_F(\lambda)^{-1}$}
The Fisher information matrix $I_F(\l)=\cov_{q_\l}(\nabla_\l\log q_\l(\t))$, where
\[\log q_\l(\t)\propto-\frac12\log|\Sigma|-\frac12(\t-\mu)'\Sigma^{-1}(\t-\mu),\]
with $\Sigma=BB'+c^2I_d$.
Let $U=U(x)$ be a matrix-valued function of a scalar $x$, and $g(U)$ a real-valued function of $U$.
Then, by the chain rule \citep{Petersen:2012}
$\nabla_x\big(g(U(x))\big)=\tr\Big[\nabla_U\big(g(U)\big)'\nabla_x\big(U(x)\big)\Big].$
Noting that, for vectors $a$ and $b$, $\nabla_U (a'U^{-1}b)=-U^{-1}ab'U^{-1}$ \citep{Petersen:2012}, we have
\beqn
\nabla_B\Big((\t-\mu)'\Sigma^{-1}(\t-\mu)\Big)=-2\Sigma^{-1}(\t-\mu)(\t-\mu)'\Sigma^{-1}B.
\eeqn
$\nabla_x(\log|U|)=\tr(U^{-1}\nabla_x U)$ because $\nabla_U(\log|U|)=U^{-1}$. Hence,
$\nabla_B\Big(\log|\Sigma|\Big)=2\Sigma^{-1}B$.
Similarly, $\nabla_c\Big(\log|\Sigma|\Big)=2c\times \tr(\Sigma^{-1})$
and $\nabla_c\Big((\t-\mu)'\Sigma^{-1}(\t-\mu)\Big)=-2c(\t-\mu)'\Sigma^{-2}(\t-\mu)$.
Hence,
\beq
\nabla_\l\log q_\l(\t)=\begin{pmatrix}
\nabla_\mu\log q_\l(\t)\\
\nabla_B\log q_\l(\t)\\
\nabla_c\log q_\l(\t)
\end{pmatrix}=\begin{pmatrix}
\Sigma^{-1}(\t-\mu)\\
-\Sigma^{-1}B+\Sigma^{-1}(\t-\mu)(\t-\mu)'\Sigma^{-1}B\\
-c\times \tr(\Sigma^{-1})+c(\t-\mu)'\Sigma^{-2}(\t-\mu)
\end{pmatrix}.
\eeq
Let $X=\Sigma^{-1}(\t-\mu)\sim\N(0,\Sigma^{-1})$. Using the results on cubic and quadratic forms of a Gaussian random vector \citep{Petersen:2012}, it can be shown that
\beq\label{eq:fact5}
I_F(\l)=\begin{pmatrix}
\Sig^{-1}&O_{d\times d}&O_{d\times 1}\\
O_{d\times d}&\Sig^{-1}BB'\Sig^{-1}+(B'\Sig^{-1}B)\Sig^{-1}&2c\Sig^{-2}B\\
O_{1\times d}&2cB'\Sig^{-2}&2c^2\tr(\Sig^{-2})
\end{pmatrix}.
\eeq 

To compute the inverse matrix $I_F(\l)^{-1}$, we first need some preliminary results. Let $A$ be a $d\times d$ matrix, $b$  a $d\times 1$ vector and $\omega$ a scalar. Then,
\[(A+bb')^{-1}=A^{-1}-\frac{1}{1+b'A^{-1}b}A^{-1}bb'A^{-1}\]
and
\[\begin{pmatrix}
A&b\\
b'&\omega
\end{pmatrix}^{-1}=\begin{pmatrix}
A^{-1}+\frac{1}{c_2}A^{-1}bb'A^{-1}&-\frac{1}{c_2}A^{-1}b\\
-\frac{1}{c_2}b'A^{-1}&\frac{1}{c_2}
\end{pmatrix},\;\;\;\;\text{ with }\;\;\;\;c_2=\omega-b'A^{-1}b.
\]
Then,
\[\Sigma^{-1}=(BB'+c^2I)^{-1}=\frac{1}{c^2}\left(I-\frac{1}{c^2+B'B}BB'\right),\]
and
\[\Sigma^{-1}B=\alpha B,\;\;\;\;\text{ with }\;\;\;\;\alpha=1/(c^2+B'B).\]
The Fisher information matrix in \eqref{eq:fact5} can written as
\[I_F(\l)=\begin{pmatrix}
\Sig^{-1}&O_{d\times d}&O_{d\times 1}\\
O_{d\times d}&A&b\\
O_{1\times d}&b'&\omega
\end{pmatrix}
\]
with
\bean
A&=&\Sig^{-1}BB^T\Sig^{-1}+(B^T\Sig^{-1}B)\Sig^{-1}=\a^2BB'+\a(B'B)\Sigma^{-1},\\
b&=&2c\Sig^{-2}B=2c\a^2B=\frac{2c}{(c^2+B'B)^2}B,\\
\omega&=&2c^2\tr(\Sig^{-2})=\frac{2}{c^2}\left[d-1+\left(\frac{c^2}{c^2+B'B}\right)^2\right].
\eean
We have
\[A^{-1}=\left[\left(1+\frac{c^2}{B'B}\right)-\frac12\left(1+\frac{c^2}{B'B}\right)^2\right]BB'+c^2\left(1+\frac{c^2}{B'B}\right)I_d,\;\;\;\;\;\text{and}\;\;\;\;A^{-1}b=\kappa B,\]
with
\[\kappa=\left[\left(1+\frac{c^2}{B'B}\right)-\frac12\left(1+\frac{c^2}{B'B}\right)^2\right]\frac{2c(B'B)}{(c^2+B'B)^2}+\frac{2c^3}{B'B(c^2+B'B)}.\]
Finally,
\[I_F(\l)^{-1}=\begin{pmatrix}
BB'+c^2I_d&O_{d\times d}&O_{d\times 1}\\
O_{d\times d}&A^{-1}+\frac{\kappa^2}{c_2}BB'&-\frac{\kappa}{c_2}B\\
O_{1\times d}&-\frac{\kappa}{c_2}B'&1/c_2
\end{pmatrix},
\]
with 
\[c_2=\frac{2}{c^2}\left[d-1+\left(\frac{c^2}{c^2+B'B}\right)^2\right]-\frac{2c\kappa B'B}{(c^2+B'B)^2}.\]

\subsubsection*{\bf Computing $\nabla_\l A(\l)$}
If we use a normal prior $p(\t)=\N(0,\s_0^2I_d)$, then
\bean
A(\l)&=&\E_{q_\l}\left(\log\frac{p(\t)}{q_\l(\t)}\right)\\
&=&-\frac{1}{2\s_0^2}\Big(\mu'\mu+\tr(BB'+c^2I)\Big)-\frac12\log|BB'+c^2I|+\text{constants}\\
&=&-\frac{1}{2\s_0^2}(\mu'\mu+B'B+dc^2)-\frac12\log\Big(c^{2d}\big(1+B'B/c^2\big)\Big)+\text{constants}.
\eean
Hence,
\beqn
\nabla_\l A(\l)=\begin{pmatrix}
\nabla_\mu A(\l)\\
\nabla_B A(\l)\\
\nabla_c A(\l)
\end{pmatrix}=\begin{pmatrix}-\frac{1}{\s_0^2}\mu\\
-\left(\frac{1}{\s_0^2}+\frac{1}{c^2+B'B}\right)B\\
-\frac{dc}{\s_0^2}-\frac1c\left(d-\frac{B'B}{c^2+B'B}\right)
\end{pmatrix}.
\eeqn

\bibliographystyle{apalike}
\bibliography{references_v1}

\end{document}